\documentclass[prl,showpacs,twocolumn,aps,superscriptaddress,a4paper]{revtex4-1}

\usepackage{pst-node}
\usepackage{amsmath}
\usepackage{graphicx}
\usepackage{latexsym}
\usepackage{amsfonts}
\usepackage{amssymb}
\usepackage{graphicx}
\usepackage{rotating}
\usepackage{bbm}
\usepackage{cancel}

\newcommand{\Tr}{ {\rm Tr} \, }
\newcommand{\be}{\begin{equation}}
\newcommand{\ee}{\end{equation}}
\newcommand{\bea}{\begin{eqnarray}} 
\newcommand{\eea}{\end{eqnarray}}
\newcommand{\nn}{\nonumber}

\begin{document}  

\def\gC{\mbox{\boldmath $C$}}
\def\gZ{\mbox{\boldmath $Z$}}
\def\gR{\mbox{\boldmath $R$}}
\def\gN{\mbox{\boldmath $N$}}
\def\gG{\mbox{\boldmath $G$}}
\def\green{\mbox{\boldmath ${\cal G}$}}
\def\grn{\mbox{${\cal G}$}}
\def\gH{\mbox{\boldmath $H$}}
\def\bA{{\bf A}}
\def\bJ{{\bf J}}
\def\bG{{\bf G}}
\def\bF{{\bf F}}
\def\bH{\mbox{\boldmath $H$}}
\def\bQ{\mbox{\boldmath $Q$}}
\def\bgS{\mbox{\boldmath $\Sigma$}}
\def\bT{\mbox{\boldmath $T$}}
\def\bU{\mbox{\boldmath $U$}}
\def\bV{\mbox{\boldmath $V$}}
\def\bgG{\mbox{\boldmath $\Gamma$}}
\def\bgL{\mbox{\boldmath $\Lambda$}}
\def\ubG{\underline{{\bf G}}}
\def\ubH{\underline{{\bf H}}}
\def\ubQ{\underline{{\bf Q}}}
\def\ubS{\underline{{\bf S}}}
\def\ubg{\underline{{\bf g}}}
\def\ubq{\underline{{\bf q}}}
\def\ubp{\underline{{\bf p}}}
\def\ubgS{\underline{{\bf \Sigma}}}
\def\bge{\mbox{\boldmath $\epsilon$}}
\def\bgD{{\bf \Delta}}

\def\bDelta{\mbox{\boldmath $\Delta$}}
\def\bcalE{\mbox{\boldmath ${\cal E}$}}
\def\bcalF{\mbox{\boldmath ${\cal F}$}}
\def\bcalG{\mbox{\boldmath $G$}}
\def\ubcalG{\mbox{\underline{\boldmath $G$}}}
\def\ubcalA{\mbox{\underline{\boldmath $A$}}}
\def\ubcalB{\mbox{\underline{\boldmath $B$}}}
\def\ubcalC{\mbox{\underline{\boldmath $C$}}}
\def\ubcalg{\mbox{\underline{\boldmath $g$}}}
\def\ubcalH{\mbox{\underline{\boldmath $H$}}}
\def\bcalK{\mbox{\boldmath $K$}}
\def\ubcalK{\mbox{\underline{\boldmath $K$}}}
\def\bcalV{\mbox{\boldmath ${\cal V}$}}
\def\ubcalV{\mbox{\underline{\boldmath $V$}}}
\def\bcalU{\mbox{\boldmath ${\cal U}$}}
\def\ubcalz{\mbox{\underline{\boldmath $z$}}}
\def\bcalQ{\mbox{\boldmath ${\cal Q}$}}
\def\ubcalQ{\mbox{\underline{\boldmath $Q$}}}
\def\ubcalP{\mbox{\underline{\boldmath $P$}}}
\def\bSS{\mbox{\boldmath $S$}}
\def\ubcalS{\mbox{\underline{\boldmath $S$}}}
\def\bff{\mbox{\boldmath $f$}}
\def\bg{\mbox{\boldmath $g$}}
\def\bh{\mbox{\boldmath $h$}}
\def\bk{\mbox{\boldmath $k$}}
\def\bq{\mbox{\boldmath $q$}}
\def\bp{\mbox{\boldmath $p$}}
\def\br{\mbox{\boldmath $r$}}
\def\bt{\mbox{\boldmath $t$}}
\def\ubh{\mbox{\underline{\boldmath $h$}}}
\def\ubt{\mbox{\underline{\boldmath $t$}}}
\def\ubk{\mbox{\underline{\boldmath $k$}}}
\def\ua{\uparrow}
\def\da{\downarrow}
\def\a{\alpha}
\def\b{\beta}
\def\g{\gamma}
\def\G{\Gamma}
\def\d{\delta}
\def\D{\Delta}
\def\e{\epsilon}
\def\ve{\varepsilon}
\def\z{\zeta}
\def\h{\eta}
\def\th{\theta}
\def\vth{\vartheta}
\def\k{\kappa}
\def\l{\lambda}
\def\L{\Lambda}
\def\m{\mu}
\def\n{\nu}
\def\x{\xi}
\def\X{\Xi}
\def\p{\pi}
\def\P{\Pi}
\def\r{\rho}
\def\bgr{\mbox{\boldmath $\rho$}}
\def\s{\sigma}
\def\us{\mbox{\underline{\boldmath $\sigma$}}}
\def\ubgm{\mbox{\underline{\boldmath $\mu$}}}
\def\S{\Sigma}
\def\ubcgS{\mbox{\underline{\boldmath $\Sigma$}}}
\def\t{\tau}
\def\f{\phi}
\def\vf{\varphi}
\def\F{\Phi}
\def\c{\chi}
\def\k{\kappa}
\def\w{\omega}
\def\W{\Omega}
\def\Q{\Psi}
\def\q{\psi}
\def\de{\partial}
\def\inf{\infty}
\def\ra{\rightarrow}
\def\bra{\langle}
\def\ket{\rangle}
\def\bbra{\langle\langle}
\def\kket{\rangle\rangle}
\def\grad{\mbox{\boldmath $\nabla$}}
\def\no{\bf 1}
\def\ze{\bf 0}
\def\uno{\underline{\bf 1}}
\def\zero{\underline{\bf 0}}

\def\dr{{\rm d}}
\def\bj{{\bf j}}
\def\br{{\bf r}}
\def\bz{\bar{z}}
\def\bart{\bar{t}}

\title{Dynamical correction to linear Kohn-Sham conductances from static 
density functional theory}

\author{S. Kurth}
\affiliation{Nano-Bio Spectroscopy Group, 
Departamento de F\'{i}sica de Materiales, 
Universidad del Pa\'{i}s Vasco UPV/EHU, Centro F\'{i}sica de Materiales 
CSIC-UPV/EHU, Avenida Tolosa 72, E-20018 San Sebasti\'{a}n, Spain} 
\affiliation{IKERBASQUE, Basque Foundation for Science, E-48011 Bilbao, Spain}
\affiliation{European Theoretical Spectroscopy Facility (ETSF)}    

\author{G. Stefanucci}
\affiliation{Dipartimento di Fisica, Universit\`{a} di Roma Tor Vergata,
Via della Ricerca Scientifica 1, 00133 Rome, Italy}
\affiliation{INFN, Laboratori Nazionali di Frascati, Via E. Fermi 40, 00044 Frascati, 
Italy}
\affiliation{European Theoretical Spectroscopy Facility (ETSF)}

\begin{abstract}
For molecules weakly coupled to leads the {\em exact} linear Kohn-Sham (KS) conductance 
can be orders of magnitude larger than the true linear conductance
due to the lack of {\em dynamical} exchange-correlation (xc) corrections. 
In this work we show how to incorporate dynamical 
effects in KS transport calculations. The only quantity needed is the {\em static} xc 
potential in the molecular junction. Our scheme provides a 
comprehensive description of Coulomb blockade without breaking the 
spin symmetry. This is explicitly demonstrated in single-wall 
nanotubes where the corrected conductance is in good agreement with 
experimental data whereas the KS conductance fails 
dramatically.
\end{abstract}

\pacs{05.60.Gg, 31.15.ee, 71.15.Mb, 73.63.-b}

\maketitle

The active field of molecular electronics \cite{cuevas} remains a challenge 
for {\em ab initio} methods. Density Functional Theory (DFT) is at present the 
only viable route for an atomistic description of complex molecular 
junctions. 
Nevertheless, DFT transport calculations still suffer from some practical 
difficulties. The fundamental sources of error in the linear conductance are the DFT 
exchange-correlation (xc) potential used to determine the Kohn-Sham 
(KS) conductance $G_{s}$ and the dynamical xc 
correction \cite{sa.2004,szvv.2005,kbe.2006,vv.2009} predicted by 
time-dependent (TD) DFT \cite{rg.1984} (see Eq.~(\ref{excond3}) below).
Assessing their importance and mutual interplay is 
especially thorny in weakly coupled molecules where level alignment 
and charging effects play a prominent role.
Toher et al. \cite{tfsb.2005} and Koentopp et al. \cite{kbe.2006} 
showed that $G_{s}$ evaluated with an accurate, and hence 
discontinuous \cite{pplb.1982}, xc 
potential is suppressed, thus capturing the Coulomb 
blockade (CB) effect at even electron numbers $N$ (closed shell). 
This may suggest the dynamical xc correction to be small. However,
at odd $N$ (open shell) $G_{s}$ can be orders of magnitude 
larger than the true conductance $G$, even when the {\em exact}   xc 
potential is employed  \cite{sk.2011}.
A satisfactory 
``DFT explanation'' of CB  is, therefore, currently missing.
In this Letter we 
provide a comprehensive picture of CB, valid for all $N$ 
without breaking the spin symmetry. The key ingredient is 
the dynamical xc correction 
which, remarkably, can be expressed exclusively in terms of {\em static}
DFT quantities. We propose a practical scheme to 
calculate $G$ and demonstrate its validity 
by comparison with recent experiments on single-wall nanotubes.

\begin{figure}[tbp]
    \begin{center}
\includegraphics[width=0.4\textwidth]{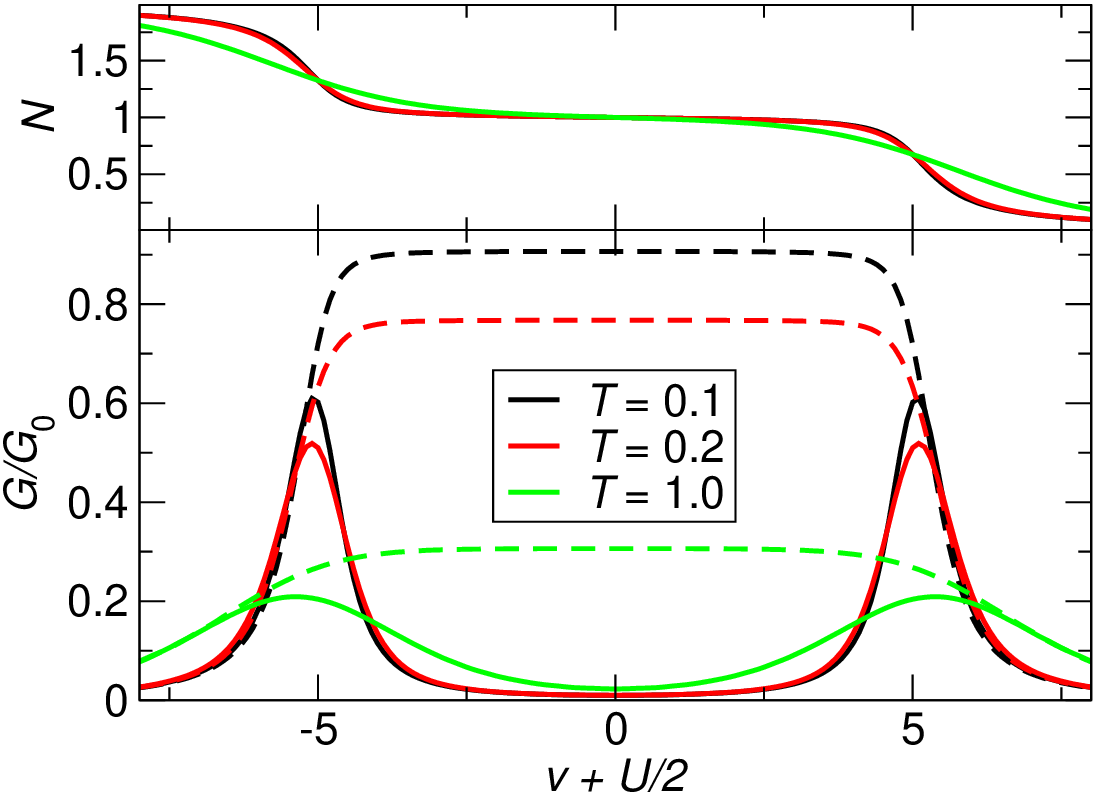}
\caption{Top: Electron number $N$ versus gate in MB and DFT 
(indistinguishable) for a single level coupled to featureless leads with 
$U=10$, $\m=0$ at various temperatures $T$ 
(all energies in units of $\g$). For these parameters 
$T_{\rm K}=\sqrt{U \g} \exp(-\frac{\pi U}{8 \g}) \simeq 0.06 $.  
Bottom: $G$ from Eq. (\ref{excond}) (solid) and
$G_{s}$ from Eq. (\ref{kscond}) (dashed) in units of
$G_{0}=2e^{2}/h$.}
\label{reveng}
\end{center}
\end{figure}

At zero temperature and when transport is dominated by a single 
resonance (open shell) $G_{s}=G$
due to the Friedel sum rule \cite{mkns.2010}. In this regime 
$G$ exhibits a Kondo plateau (no CB peaks) and the discontinuity 
is essential for $G_{s}$ to reproduce the plateau \cite{sk.2011,blb.2012,tse.2012}.
At temperatures higher than the Kondo temperature $T_{\rm K}$, 
the exact discontinuous xc potential gives instead a $G_{s}\gg G$ \cite{sk.2011}.
To understand this discrepancy we model the resonance with 
a single level (HOMO or LUMO) of energy $v$ and Coulomb repulsion  
$U$ coupled to left ($L$) and right ($R$) featureless leads 
contributing $\g=\g_{L}+\g_{R}$ to the 
broadening of the spectral peaks. 
Given the Many-Body (MB) spectral function $A(\w)$
the number of electrons is
\be
N=2\int f(\w)A(\w),\quad\quad \int\equiv \int\frac{d\w}{2\p}
\label{exdens}
\ee
whereas the linear (zero-bias) conductance reads
\be
G=-2\frac{\g_{L}\g_{R}}{\g}\int f'(\w)A(\w)
\label{excond}
\ee
with the Fermi function $f(\w)=1/(e^{\b(\w-\m)}+1)$  at inverse 
temperature $\b=1/T$ and chemical potential $\m$.
At $T\gg T_{\rm K}$
the Abrikosov-Suhl (AS) resonance is suppressed and 
the spectral function is well represented by \cite{hjbook}
\be
A(\w)=n\,L_{\g}(\w-v-U)+(1-n)\,L_{\g}(\w-v)
\label{moda}
\ee
where $n=N/2$ and 
$L_{\g}(\w)=\g/(\w^{2}+\g^{2}/4)$. 
For the KS system the spectral function is
$A_{s}(\w)=L_{\g}(\w-v-v_{\rm Hxc}[N])$.
The Hartree-xc (Hxc) potential $v_{\rm Hxc}$ is such that the number of 
electrons $N$ which solves
$N=2\int f(\w)A_{s}(\w)$
is the same as in Eq. (\ref{exdens}). We obtain 
$v_{\rm Hxc}$ by reverse engineering and find that, as function of $N$, 
it has the shape of a smeared step function (not shown). The smearing 
is due to the level broadening $\gamma$ induced by contacting the level 
to the leads and develops into a true discontinuity (at $N=1$) only in the 
limit $\gamma \to 0$, as it should \cite{blb.2012,tse.2012}. 
This $v_{\rm Hxc}$ is then used to calculate the KS conductance from
\be
G_{s}=-2\frac{\g_{L}\g_{R}}{\g}\int f'(\w)A_{s}(\w).
\label{kscond}
\ee
Despite the fact that the MB and DFT $N$-$v$ curves are {\em identical}, 
see Fig.~\ref{reveng} top, 
the CB peaks present in $G$ are completely absent in $G_s$, see 
Fig.~\ref{reveng} bottom. The physical 
situation discussed here is distinct from that of Ref. 
\onlinecite{tfsb.2005} where the discontinuity keeps the HOMO doubly 
occupied and the LUMO empty when gating the molecule 
(closed shell). The discontinuity correctly suppresses $G_{s}$ at even $N$ 
but has the opposite effect at odd $N$.

{\em Dynamical xc effects}: 
Open-shell molecules in the CB regime 
are probably the most striking example of the inadequacy of 
standard DFT transport calculations. Below we  derive an exact formula for 
$G$ in terms of TDDFT 
quantities. We  
take the leads as two jellia
(the argument can be generalized to  more realistic leads) and choose
$z$ as the longitudinal coordinate so that $z\ra-\inf$ 
is in the left lead, $\a=L$, whereas  $z\ra \inf$ 
is in the right lead, $\a=R$. 
Let $\d V^{\a}$ be the variation in the classical potential (external 
plus Hartree) of lead 
$\a$. 
This perturbation generates a current \cite{sa.2004,kbe.2006}
\be
\d I=(\d V^{R}-\d V^{L}+\d V^{R}_{\rm xc}-\d V^{L}_{\rm xc})G_{s}
\label{iks}
\ee
where $\d V^{\a}_{\rm xc}=\lim_{t\ra\inf}\lim_{z\ra s_{\a}\inf} \d v_{\rm 
xc}({\bf r},t)$, $s_{R/L}=\pm$, 
is the asymptotic value of the variation of the xc potential $\d 
v_{\rm xc}$. From linear-response TDDFT 
\be
\d V^{\a}_{\rm xc}=\int 
dt'd{\bf r}' \lim_{t\ra\inf}\lim_{z\ra s_{\a}\inf} 
f_{\rm xc}({\bf r},{\bf r}';t-t')\d n({\bf r}',t')
\label{dv=fxcdn}
\ee
where $f_{\rm xc}$ is the TDDFT kernel and $\d n({\bf r},t)$ is the 
density variation. 
The assumption of a steady state implies that 
$f_{\rm xc}\ra 0$ for $|t-t'|\ra\inf$ and 
$\d n({\bf r},t\ra\inf)=s_{\a}\d n$ for ${\bf r}$ in lead $\a$. In 
Eq.~(\ref{dv=fxcdn}) the contribution of the molecular region 
to the spatial integral  is negligible in the thermodynamic 
limit. If we define
\be
f_{\rm xc}^{\a\b}=\int dt'
\int_{\rm lead\;\b} d{\bf r}'\lim_{z\ra s_{\a}\inf}
f_{\rm xc}({\bf r},{\bf r}';t')
\ee
then $\d V^{\a}_{\rm xc}=\sum_{\b=L,R}f_{\rm xc}^{\a\b}s_{\b}\d n$.
We emphasize that $f_{\rm xc}^{\a\b}$ is not 
the static DFT kernel since the limit $t\ra \inf$ is taken {\em after} 
the limit $|z|\ra \inf$ and these two limits, in general, do not 
commute \cite{note}. 
This implies that we cannot model $f_{\rm xc}^{\a\b}$ by performing 
DFT calculations on leads of finite length.
Inserting the expression for $\d V^{\a}_{\rm xc}$ into Eq. (\ref{iks}) we find
$\d I=(\d V^{L}-\d V^{R})G_{s}-\F G_{s}\d n$
where
\be
\F\equiv f_{\rm xc}^{RL}+f_{\rm xc}^{LR}
-f_{\rm xc}^{RR}-f_{\rm xc}^{LL}.
\ee
The expression for $\d I$ is correctly gauge invariant. The 
kernel $f_{\rm xc}$ is defined up to the addition of an arbitrary function 
$q({\bf r})+q({\bf r}')$ \cite{hg.2012} and $\F$ 
is invariant under this transformation.
In conclusion
\be
G\equiv \frac{\d I}{(\d V^{R}-\d V^{L})}=\frac{G_{s}}{1+\chi\F G_{s}}.
\label{excond3}
\ee
The quantity $\chi\equiv \d n/\d I\simeq 1/(v_{F}\s)$ 
with $v_{F}$ the Fermi 
velocity and $\s$ the cross section of the leads \cite{uksskvlg.2012}.
In the following we define $\chi\F G_{s}$ as the 
{\em dynamical} xc correction since $\F$ is expressed in terms 
of the TDDFT kernel.

{\em Approximations to $\F$}:
To gain some insight into the density dependence of $\chi\F$ we consider 
again the single level model. For $N\neq 1$ and $T<\gamma$ 
the real and KS systems respond similarly and 
consequently $G\simeq G_{s}$. On the 
other hand for $N=1$ we have $G\simeq 0$ whereas $G_{s}\simeq 
G_{0}=2e^{2}/h$ the quantum of conductance.
Therefore $\chi\F$ is small for $N\neq 1$ and large for $N=1$. 
Interestingly the quantity $\de  v_{\rm Hxc}/\de N$ behaves similarly.
Is there any relation between $\chi\F$ and 
$\de  v_{\rm Hxc}/\de N$? If so this relation would simplify enormously the 
problem of estimating the dynamical xc correction since $\de v_{\rm 
Hxc}/\de N$ can be calculated from static DFT. In the following we 
show that in the CB regime this 
relation does actually exist.

Consider the system in equilibrium.
Using Eq. (\ref{exdens}) the
compressibility $\kappa=\de N/\de \m$ can be written as
$\kappa=\frac{\g}{\g_{L}\g_{R}}G+
2\int f(\w)\de A(\w)/\de \m$,
where we identified the conductance $G$ of Eq.~(\ref{excond}).
If we define the quantity 
$R\equiv-2\int f(\w)\de A(\w)/\de N$
then $\kappa=\frac{\g}{\g_{L}\g_{R}}G/(1+R)$.
As the MB and DFT densities are the same, the MB and 
DFT compressibilities are the same too. Hence
$\kappa=
\frac{\g}{\g_{L}\g_{R}}G_{s}+2\int f(\w)\de 
A_{s}(\w)/\de \m$,
where we identified the KS conductance $G_{s}$ of Eq.~(\ref{kscond}).
The KS spectral function depends on $\m$  
through  $N$, and the dependence on $N$ is all 
contained in $v_{\rm Hxc}$. Since 
$\frac{\de A_s}{\de v_{\rm Hxc}} = - \frac{\de A_{s}}{\de\w}$ (see definition 
of $A_s$ below Eq.~(\ref{moda})) we have 
$\frac{\de A_{s}}{\de\m}=-\frac{\de A_{s}}{\de\w}\frac{\de v_{\rm Hxc}}{\de N}\frac{\de 
N}{\de\m}$. Using this result under the integral, solving for 
$\kappa$ and equating the MB and DFT expressions one easily obtains 
\be
\frac{G}{G_{s}}=\frac{1+R}{1+\frac{\g}{\g_{L}\g_{R}}G_{s}\frac{\de 
v_{\rm Hxc}}{\de N}}.
\label{GT}
\ee
No approximations have been made so far. 
Let us study the dependence 
of $R$ on temperature. 

We first consider the low temperature case.
For simplicity we take $\g_{L}=\g_{R}$ and set $v=-U/2$ 
at the particle-hole (ph) symmetric point (half-filling). At zero temperature $G=G_{s}=G_{0}$ 
and hence 
$R=R_{0}\equiv \frac{4G_{0}}{\g}\frac{\de v_{\rm Hxc}}{\de N}$ \cite{check}.
For 
temperatures $T>T_{\rm K}$ the AS resonance broadens and 
its height decreases as $h(T/T_{\rm K})$ where $h$ is a universal 
function which approaches zero at high $T$ \cite{c.2000}. This means that 
$R\simeq h(T/T_{K})R_{0}$ remains large until the 
AS resonance disappears. No simple relation between $\F$ 
and $\de v_{\rm Hxc}/\de N$ exists when Kondo correlations are 
present. 

\begin{figure}[tbp]
    \begin{center}
\includegraphics[width=0.45\textwidth]{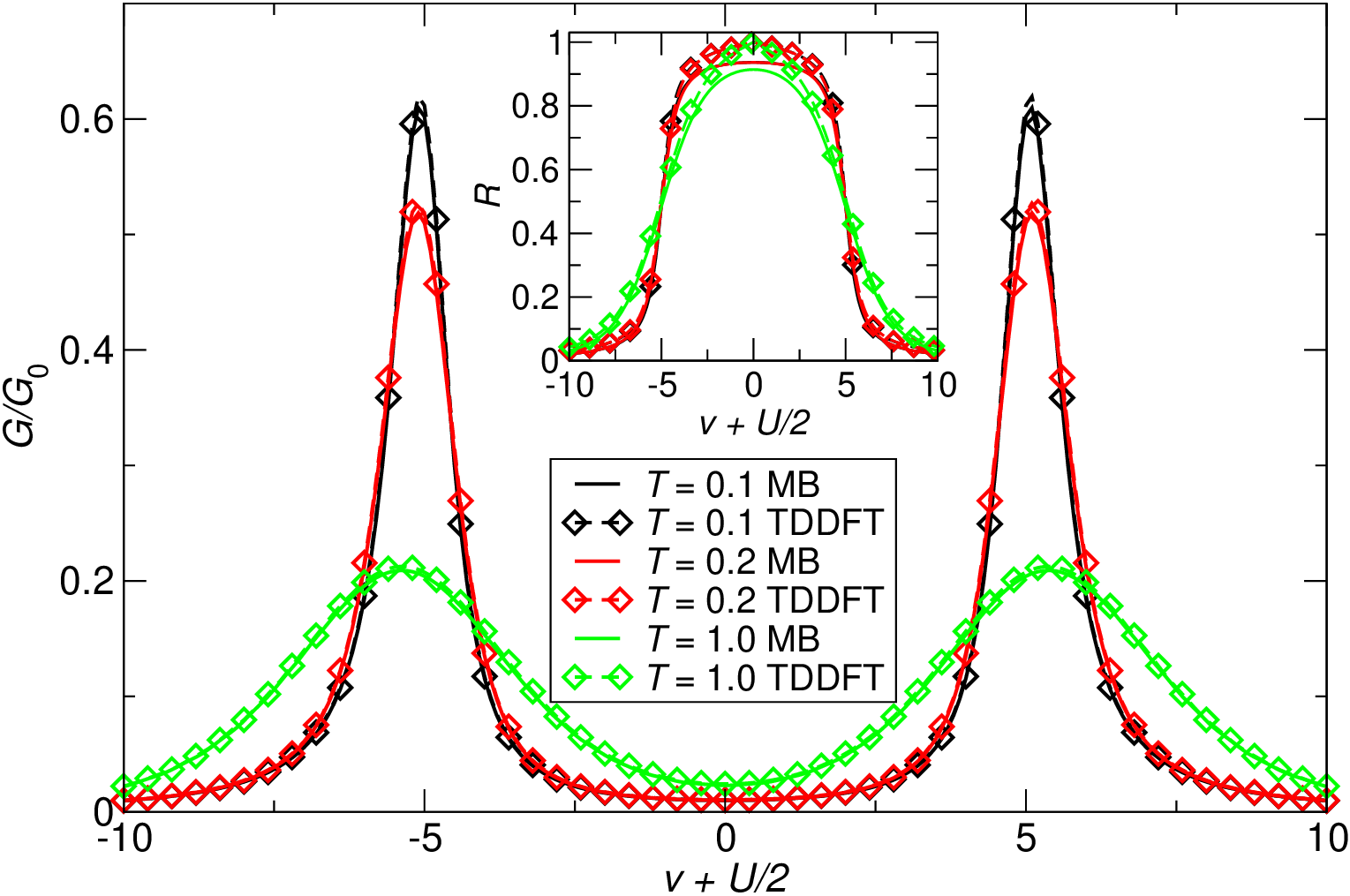}	
\caption{Linear conductance from Eq.~(\ref{excond}) 
using the spectral function Eq.~(\ref{moda}) (MB, solid) and from 
Eq.~(\ref{GT2}) (TDDFT, dashed).
The inset shows a comparison between the exact and the 
approximate  $R$. Same parameters as in Fig. \ref{reveng}.}
\label{Gvv}
\end{center}
\end{figure}

At temperatures 
$T\gg T_{\rm K}$  thermal 
fluctuations destroy the Kondo effect and the MB spectral 
function is well approximated by Eq.~(\ref{moda}). 
Therefore
$R(v)=I(v)-I(v+U)$ where $I(E)\equiv \int f(\w)L_{\g}(\w-E)$.
We can derive a more convenient expression for $R$ 
by inserting Eq. (\ref{moda}) into Eq. 
(\ref{exdens}) to find
\be
N=\frac{2 I(v)}{1+I(v)-I(v+U)},
\ee
and hence
$1+R=2I(v)/N$. Unfortunately $I(v)$  is not an explicit function of $N$ 
due to the implicit dependence of $v=v[N]$.
However, for $v<\mu$, or equivalently for $N<1$, we have $I(v+U)\ll 1$.  
Thus for $N<1$ we can write 
$N\simeq 2I(v)/(1+I(v))$ from which we infer 
$I(v)\simeq N/(2-N)$. Using ph symmetry we therefore approximate
$R$ by the explicit function 
$1+R= 2/(1+|\d N|)$
where $\d N=N-1$. 
Inserting this into Eq. (\ref{GT}) we deduce the main result of this 
Letter
\be
\frac{G}{G_{s}}= \frac{2}{1+|\d N|}
\frac{1}{1+\frac{\g}{\g_{L}\g_{R}}G_{s}\frac{\de v_{\rm Hxc}}{\de N}}.
\label{GT2}
\ee
Equation (\ref{GT2})
provides a simple and implementable formula to correct the KS conductance. 
In fact, the dynamical xc correction of Eq.~(\ref{excond3}) is entirely
expressed in terms of {\em static} DFT quantities. Moreover, whereas 
$\F$ involves the TDDFT kernel with coordinates in the {\em leads} the 
correction in Eq.~(\ref{GT2}) involves only the DFT $v_{\rm Hxc}$ 
in the {\em molecular junction}. 
The accuracy of Eq. (\ref{GT2}) is examined in Fig. \ref{Gvv}, 
and benchmarked 
against the MB conductance of Eq. (\ref{excond}).
Even though the approximate $R$ is 
not on top of the exact one, see inset, the agreement between the
two conductances is extremely good. The position, width and 
height of the peaks as well as the decay for large $|v|$ are all well 
reproduced. Most importantly the plateau of $G_{s}$, see Fig. 
\ref{reveng}, is completely gone.

{\em Application to physical systems}: In real molecules $v_{\rm 
Hxc}$ is a $\br$-dependent functional of the density. We write 
$v_{\rm Hxc}({\bf r})=
\d v_{\rm Hxc}({\bf r})+\bar{v}_{\rm Hxc}$ as the sum of  
a functional $\d v_{\rm Hxc}$ with a
weak dependence on $N=\int_V d{\bf r}\,n({\bf r})$ and a spatially uniform part 
$\bar{v}_{\rm Hxc}=\frac{1}{V}\int_{V}d\br\, v_{\rm Hxc}({\bf 
r})$, where the integral is over the volume $V$ of the molecular 
junction. For weakly coupled molecules $\bar{v}_{\rm Hxc}$ exhibits 
sharp steps as function of $N$ when $N$ crosses an integer.
If we are at resonance and  spin fluctuations 
are suppressed (no Kondo effect) then the KS conductance 
must be corrected according to Eq. (\ref{GT2}) in which $\de v_{\rm 
Hxc}/\de N\ra \de \bar{v}_{\rm xc}/\de N$. 
Below we argue that 
this correction applies out-of-resonance too.
Let $\m$ be in the HOMO-LUMO gap and consider 
a two-level system with $\G_{\a}$ the $2\times 2$ broadening matrix.
For general $\G_{\a}$ no simple analytic relation between $G$ and   
$N$ exists. However if $\G_{\a,ml}=(\g_{\a}/2)\d_{ml}$
then $N=2\int f(\w)\Tr[A(\w)]$ and 
$G=-2\frac{\g_{L}\g_{R}}{\g}\int f'(\w)\Tr[A(\w)]$. 
Discarding the dependence of $\d v_{\rm Hxc}$ on $\m$ (which is weak 
by definition) 
we can go through the same steps of the single-level derivation and 
find again Eq. (\ref{GT2}). It is therefore reasonable to expect that 
the KS conductance should be corrected even 
out-of-resonance (closed shell)  and that this correction should be proportional to 
$G_{s}\de \bar{v}_{\rm Hxc}/\de N$. 
 
We here propose a practical scheme to calculate $G$ from 
DFT. Given the KS Hamiltonian matrix $h_{{\rm KS},ml}=\d_{ml}\e_{l}$ 
and the broadening matrices $\G_{\a,ml}$ we determine
the density and $G_{s}$ in the usual manner. $G$ is then 
obtained from Eq. (\ref{GT2}) where $\d N$ is the deviation of 
$(N-{\rm Int}[N])$ from $1$ 
whereas $\g_{\a}=\g_{\a}(N)= \G_{\a,HH}$ if $\m\simeq 
\e_{H}$ (resonance, open shell) and $\g_{\a}(N)= 
\frac{1}{2}(\G_{\a,HH}+\G_{\a,LL})$ if $\m\simeq 
\frac{1}{2}(\e_{L}+\e_{H})$ 
is in the HOMO-LUMO gap (out-of-resonance, closed shell).
One could improve the approximation to $\g_{\a}$ using different 
weights, but the qualitative features of the results are independent of these 
details.

To appreciate the decisive impact of the dynamical xc 
correction we consider two paradigmatic junctions in which 
$\d v_{\rm Hxc}$ can be discarded. 
For $\bar{v}_{\rm Hxc}$ we choose a best fit of the zero-temperature 
limit of the single-level Hxc potential, but now 
sum over all possible charged states of the molecule \cite{psc.2012}, i.e., 
\be
\bar{v}_{\rm Hxc}=
\sum_{K}\frac{U(K)}{\p}\arctan\left(\frac{N-K}{W(K)}\right).
\label{hxcpotmod}
\ee
The charging energies $U(N)$ are given by the xc part of the 
derivative discontinuity of the 
molecule with $N$ electrons \cite{pplb.1982}. 
For the widths we take $W(N)= 0.16\, \g(N)/U(N)$ which is consistent with 
Ref.~\onlinecite{es.2011}
 \begin{figure}[tbp]
    \begin{center}
\includegraphics[width=0.48\textwidth]{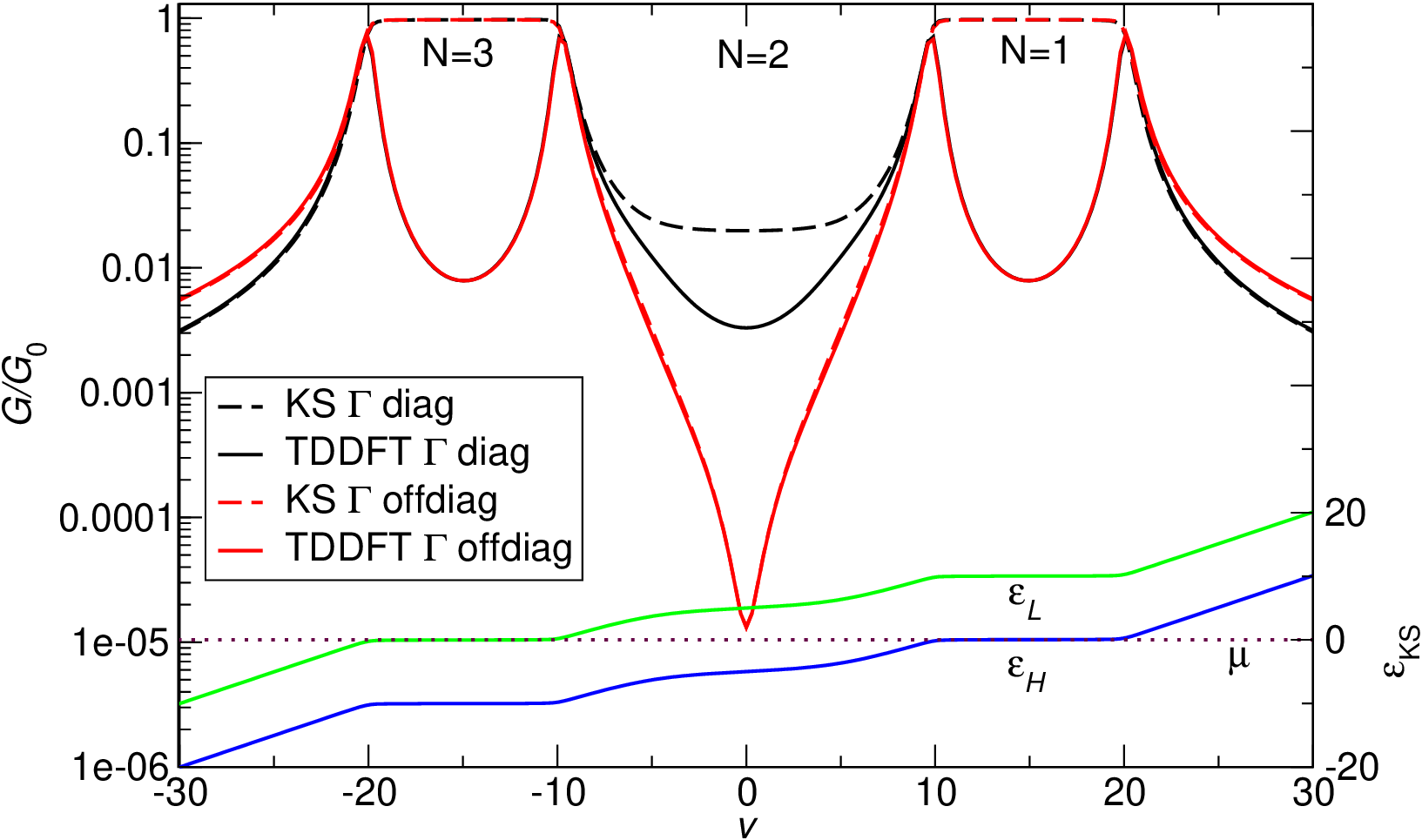}	
\caption{Linear conductance from Eq.~(\ref{kscond}) (KS, dashed) and from 
Eq.~(\ref{GT2}) (TDDFT, solid)
for the HOMO-LUMO model with diagonal and off-diagonal $\G$-matrices 
(left axis) and KS energies 
$\e_{H/L}=\e_{0H/L}+\bar{v}_{\rm Hxc}$ (right axis). The electron number 
$N$ for different ranges of $v$ is also indicated.
}
\label{G-homo-lumo}
\end{center}
\end{figure}

{\em HOMO-LUMO model}: We study a two-level system with 2 
electrons in the HOMO in the charge neutral state. Let 
$\e_{0H}=-\e_{0L}=-\e_{0}<0$ be the noninteracting single-particle energies, 
\begin{equation*}
\G_{L}=\G_{R}=\frac{\g}{2}\left(
\begin{array}{cc} 1 & 1 \\ 1 & 1 
\end{array}
\right)
\end{equation*}
and $U(N)=U$ independent of $N$. We solve the self-consistent equation 
for the density with $U=10$, $\e_{0}=5$, $\m=0$ and $\b=10$ (all energies 
in units of $\g$). $G_{s}$ and $G$ from Eq. (\ref{GT2}) are shown in Fig. 
\ref{G-homo-lumo} (left axis). As expected the discontinuity of $v_{\rm Hxc}$  
opens a gap in $G_{s}$ for even $N$, in agreement with 
the results of Ref. \onlinecite{tfsb.2005}. Here the 
dynamical xc correction only weakly affects $G_{s}$  since 
$\de \bar{v}_{\rm Hxc}/\de N$ is multiplied by $G_{s}\ll 1$.
For odd $N$ 
the KS conductance exhibits a Kondo plateau due to the pinning of 
the KS level to $\m$, see right axis. This is the regime 
previously discussed and no CB is observed. 
The dynamical xc correction remedies this serious deficiency by 
correctly suppressing the plateau. 
The results remain essentially unaltered if the 
off-diagonal matrix elements of $\G_{\a}$ are discarded 
\cite{interference}.

\begin{figure}[tbp]
    \begin{center}
\includegraphics[width=0.47\textwidth]{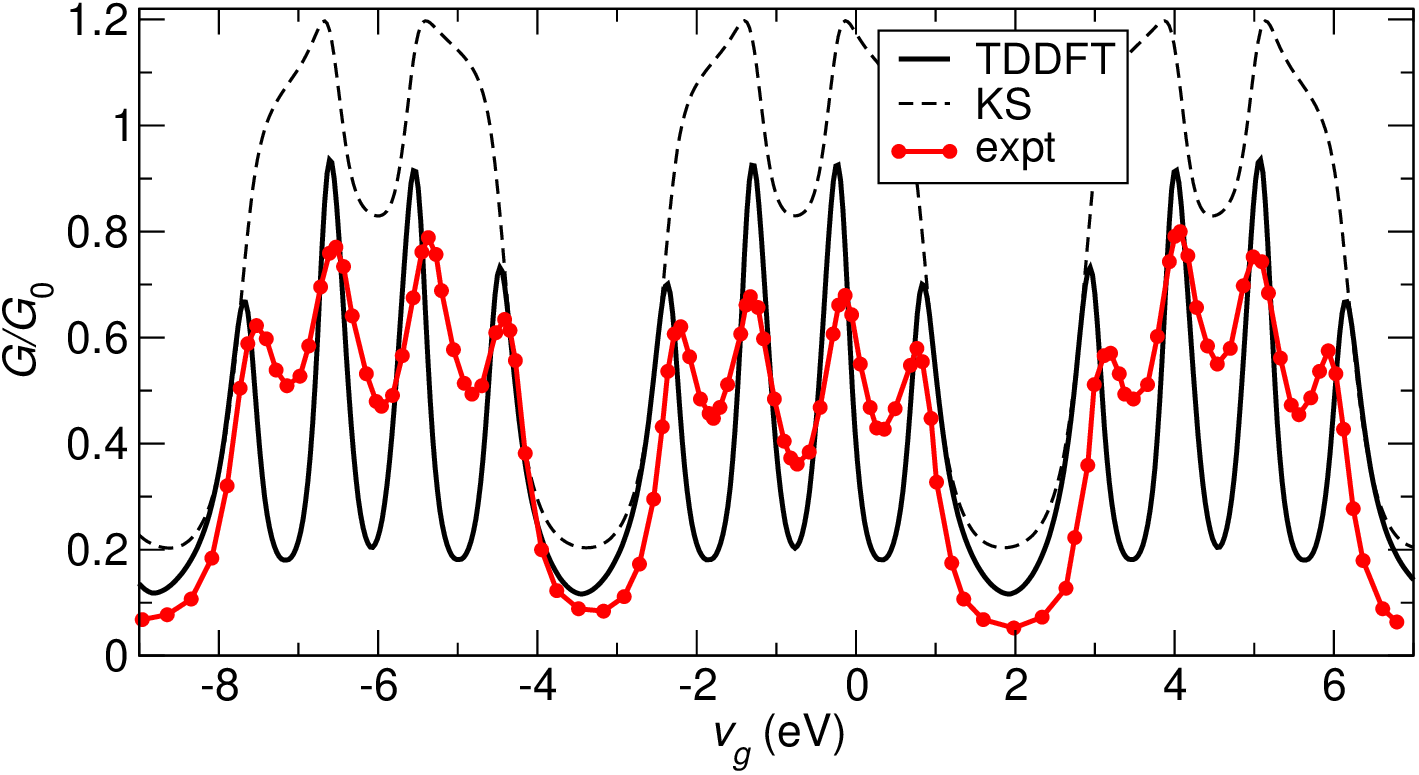}	
\caption{Linear KS and TDDFT conductance (Eq.~(\ref{GT2})) 
for a SWNT quantum dot in comparison to 
experimental conductance from Ref. \onlinecite{lbp.2002}, 
as function of gate voltage $v_{g}$.}
\label{SWNT-comparison}
\end{center}
\end{figure}

{\em SWNT}: Experimental evidence of CB oscillations  
has recently been reported in metallic single-wall nanotubes (SWNT) 
quantum dots \cite{lbp.2002,bbnis,sjkdkz.2005}. We now analyze the 
performance of Eq.~(\ref{GT2}) in these systems.
The finite length of the SWNT causes a 
level quantization of the twofold degenerate bands. 
Since the wavevector is a good quantum number our approximation $\d 
v_{\rm Hxc}=0$ is justified.
For a SWNT quantum dot the constant interaction model \cite{kat.2001}
has been refined by Oreg {\em et al.} \cite{obh.2000} to account for 
the observed fourfold periodicity in the electron addition energy. 
We constructed the KS Hamiltonian corresponding to this model and 
approximated the broadening matrix as 
$\G_{\a,ml}=(\g/2)\d_{ml}$ \cite{suppl} 
(no visible interference \cite{interference} from 
experiment). In Fig. \ref{SWNT-comparison} we compare the KS, 
TDDFT and experimental conductance versus the gate voltage $v_{g}$. 
We clearly see that the conductance of Eq.~(\ref{GT2}) 
correctly exhibits the fourfold periodicity and 
represents a considerable improvement 
over $G_{s}$ which, instead, shows two deformed Kondo plateaus per period. 
The qualitative 
behavior of $G$ and $G_{s}$ does not change by varying the parameters 
within a reasonable range around the average values reported in Ref. 
\onlinecite{lbp.2002}. 

In conclusion we proposed a practical scheme to correct  KS 
conductances. We highlighted the role of the discontinuity not only for 
an accurate $G_{s}$ but also for an accurate dynamical xc correction to $G$. 
Approaches to generate discontinuous xc potentials are 
emerging both in the static \cite{discpot} and dynamical \cite{tddft_xc} case. 
Our theory provides a coherent picture of  CB  within (TD)DFT without 
breaking the spin symmetry. By application to two different  molecular 
junctions we further showed that the dynamical 
xc correction always reduces $G_{s}$, thus contributing
to close the gap between theoretical predictions and 
experimental measurements.

S.K. acknowledges funding by the ``Grupos Consolidados UPV/EHU del
Gobierno Vasco'' grant No. IT578-13. 
G.S.  acknowledges funding by MIUR FIRB 
grant No. RBFR12SW0J.
We acknowledge support through 
travel grants (Psi-K2 4665 and 3962 (G.S.), and Psi-K2 5332 (S.K.)) of the 
European Science Foundation (ESF).

\newpage
\begin{widetext}

\begin{center}
{\Large{\bf Supplemental Material}}
\end{center}\vspace{0.6cm}
In this Supplemental Material we refer to the paper as I. The many-body 
Hamiltonian of the SWNT quantum dot was proposed by Oreg et al. (PRL {\bf 85}, 
365 (2000)) and reads
\be
H=\sum_{l\n\s}\e_{0l\n}n_{l\n\s}+\frac{1}{2}E_{C}\left(N-N_{0}\right)^{2}
+\d U\sum_{l\n}n_{l\n\ua}n_{l\n\da}+J N_{\ua}N_{\da}.
\label{oregham}
\ee
Here $\s$ is the spin index, $\n=0,1$ is the band index and $l$ is the 
integer of the quantized quasi-momentum of the electrons. The Hamiltonian is 
written in terms of the occupation numbers $n_{l\n\s}$ since 
$N_{\s}\equiv \sum_{l\n}n_{l\n\s}$ (total number of electrons 
with spin $\s$) and $N=N_{\ua}+N_{\da}$ (total number of electrons). The 
finite length of the SWNT causes a finite subband mismatch $\d$ so that the  
single-particle energies are
\be
\e_{0l\n}=\left\{
\begin{array}{ll} 
l\D-\d &\quad{\rm for}\quad \n=0 \\
l\D&\quad{\rm for}\quad \n=1
\end{array}
\right.
\ee
with $\D$ the average level spacing. In Eq.~(\ref{oregham}) the parameter 
$E_{C}$ is the charging energy ($N_{0}$ is the number of electrons of the 
charge neutral SWNT quantum dot), $\d U$ is the extra charging energy for 
two electrons in the same energy level and $J$ is the exchange energy 
between electrons of opposite spin. With the parameters of W. Liang et al. 
(PRL {\bf 88}, 126801 (2002)) when an extra electron  enters the nanotube it 
occupies the lowest available single-particle energy level.Thus the 
{\em aufbau} is the same as that of the noninteracting Hamiltonian. 
Accordingly the spin of the ground state is $0$, $1/2$, $0$, $1/2$, \ldots 
for 0, 1, 2, 3,  \ldots extra electrons. Let $E(N)$ be the energy of the SWNT 
with $N$ extra electrons. Using Eq.~(\ref{oregham}) it is straightforward 
to obtain
\bea
E(0)&=&0,\nn\\
E(1)&=&\D-\d+\frac{1}{2}E_{C},
\nn\\E(2)&=&2\D-2\d+\frac{1}{2}4E_{C}+\d U +J,
\nn\\
E(3)&=&3\D-2\d+\frac{1}{2}9E_{C}+\d U+2J,
\eea
and so on. The function $E(N)$ with $N$ a real continuous variable has a 
discontinuous derivative at integers $N$ and the size of this discontinuity  
is given by $\D(N)=E(N+1)-2E(N)+E(N-1)$, see 
Perdew et al. (PRL {\bf 49}, 1691 (1982)). One finds
\bea
\D(1)&=&E_{C}+\d U +J,\nn\\
\D(2)&=&\d+E_{C}-\d U ,
\label{delta}
\eea
and $\D(N)=\D(K)$ if $N-K=0$ mod2 with $K=1,2$. We now turn to the KS system. 
The KS Hamiltonian of the SWNT has the general form
\be
H_{\rm KS}=\sum_{l\n\s}(\e_{0l\n}+v_{{\rm Hxc},l\n}[\{n\}])\,n_{l\n\s}
\ee
The Hxc potential $v_{{\rm Hxc},l\n}$ has the property that for a given 
chemical potential the ground state occupations $n_{l\n\s}$ of $H$ and of 
$H_{\rm KS}$ are identical. Since $H$ is diagonal in the occupation basis and 
follows the {\em aufbau} of the noninteracting Hamiltonian, we have that 
$v_{{\rm Hxc},l\n}[\{n\}])=\bar{v}_{\rm Hxc}[N]$ is uniform and depends only on 
the total number of particles $N$. Thus, for the SWNT the fluctuation 
$\d v_{\rm Hxc}$ around the average $\bar{v}_{\rm Hxc}$ (these quantities have 
been introduced in ``Application to physical systems'' of I) is exactly zero. 
For an isolated SWNT the potential $\bar{v}_{\rm Hxc}[N]$ is discontinuous 
every time $N$ crosses an integer and the size of the step, $U$, is given by 
the xc part of the discontinuity $\D$. From Eqs.~(\ref{delta}) we find 
\bea
U(1)&=&E_{C}+\d U +J,\nn\\
U(2)&=&E_{C}-\d U ,
\label{u}
\eea
and $U(N)=U(K)$ if $N-K=0$ mod2 with $K=1,2$. The average values of these 
parameters can be found in W. Liang et al. (PRL {\bf 88}, 126801 (2002)). 
Here we change them slightly to match the peak positions of the SWNT of 
length $\simeq 100$ nm (all energies are in meV): $\D=9.2$, $\d=2.27$,  
charging energy $E_{C}=2.485$, exchange energy $J=0.7$, extra charging 
energy for doubly occupied levels $\d U=0.37$. The sharp steps of 
$\bar{v}_{\rm Hxc}$ are smeared when the SWNT is brought in contact with 
the leads. To find the functional form of $\bar{v}_{\rm Hxc}$ we observe that 
the exact smeared potential which produces the electron number of Fig. 1 of I 
(single level model) is well approximated by 
\be
v_{\rm Hxc}[N]=\frac{U}{2}+\frac{U}{\p}\arctan(\frac{N-1}{W})
\ee
where $W\simeq 0.16 \g/U$. Therefore a good approximation to the Hxc 
potential is [cf. with Eq.~(13) in I]
\be
\bar{v}_{\rm Hxc}[N]=\sum_{K}\frac{U(K)}{\p}\arctan(\frac{N-K}{W(K)})
\label{vhxc}
\ee
where the $U$'s are given in Eq.~(\ref{u}). In Eq.~(\ref{vhxc}) we have 
chosen the smearing parameter $W(K)=0.16 \g(K)/U(K)$ where $\g(K)=\g$ is 
independent of $K$ since the broadening of the Coulomb blockade peaks 
observed in W. Liang et al. (PRL {\bf 88}, 126801 (2002)) is approximately the 
same for all peaks (see below).  We study the addition and removal of up to 
6 electrons from the charge neutral SWNT due to changes of the gate $v$. In 
order to avoid finite-size effects we considered 22 consecutive 
single-particle levels $\e_{0l\n}$. This ensures that for $16 \leq N \leq 28$ 
($N_0=22$) the KS and TDDFT conductances are converged. The broadening of 
these levels (due to the contacts) is described by the broadening (or 
hybridization) matrix
\be
[\G_{\a}]_{l\n,l'\n'}=2\p \sum_{k}T_{k\a,l\n}\d(\w-\e_{k\a})T_{k\a,l'\n'}
\ee
where $T_{k\a,l\n}=\bra\f_{k\a}|H_{\rm tot}|\f_{l\n}\ket$ is the matrix element 
of the lead-SWNT-lead Hamiltonian $H_{\rm tot}$ between a bulk eigenstate 
$\f_{k\a}$ with energy $\e_{k\a}$  of lead $\a=L,R$ and the eigenstate 
$\f_{l\n}$ of the SWNT. We discard the off-diagonal matrix elements of the 
broadening matrix since in the CB regime transport is dominated by a single 
resonance and interference effects (accounted for by the off-diagonal part 
of $\G_{\a}$) can be neglected. Furthemore, we take the diagonal matrix 
elements of $\G_{\a}$ all the same, which is consistent with the observation 
below Eq.~(\ref{vhxc}). Thus
\be
[\G_{\a}]_{l\n,l'\n'}=\d_{ll'}\d_{\n\n'}\g_{\a}
\ee
with $\g_{L}=\g_{R}\equiv\g/2$. We solve the self-consistent equation
\be 
N=2\int \frac{d\w}{2\p}f(\w)\Tr[A_{s}(\w)]
\ee 
with the KS spectral function 
\be
[A_{s}(\w)]_{l\n,l'\n'}=\d_{ll'}\d_{\n\n'}\frac{\g}{(\w-\e_{0l\n}-
\bar{v}_{\rm Hxc}[N]-v)^{2}+\g^{2}/4}
\ee
for temperatures $T<\g$. 
\begin{figure}[tbp]    
\begin{center}
\includegraphics[width=0.8\textwidth]{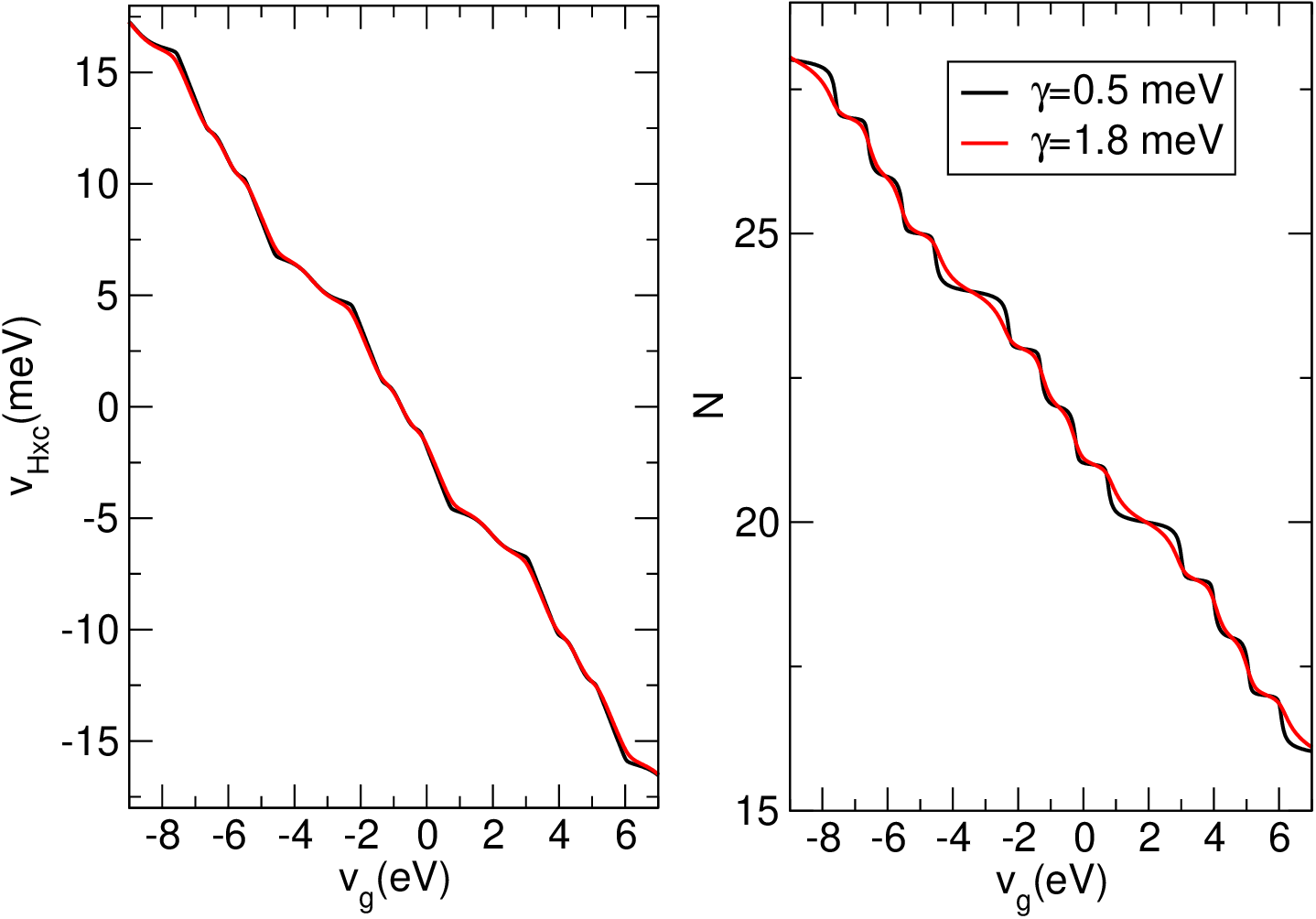}	
\caption{KS Hxc potential (Eq.~(\ref{vhxc})) and number of electrons on 
the SWNT quantum dot for different values of the coupling $\gamma$ to 
the leads.}
\label{suppl_vhxc_n}
\end{center}
\end{figure}
In Fig.~\ref{suppl_vhxc_n} we plot $\bar{v}_{\rm Hxc}[N]$ as a function of 
$N$ as well as the self-consistent value of $N$ as a function of 
$v_{g}=v_{0}+\a v$ for two different values of $\gamma$.  Here $v_{0}$ is 
chosen in order to have the same reference energy as in 
W. Liang et al. (PRL {\bf 88}, 126801 (2002)) and we 
estimated the ratio $\a=C/C_{g}\simeq 250$ between the total capacitance and 
the gate capacitance from the experimental data. For a given $v$ we 
calculate $A_{s}$ with $N=N[v]$ and then the KS conductance from 
\be
G_{s}=-\frac{\g}{2}\int\frac{d\w}{2\p}f'(\w)\Tr[A_{s}(\w)].
\ee
Subsequently we calculate $\de\bar{v}_{\rm Hxc}/\de N$ and correct 
$G_{s}$ according to 
\be
\frac{G}{G_{s}}= \frac{2}{1+|\d N|}\frac{1}{1+\frac{\g}{\g_{L}\g_{R}}G_{s}
\frac{\de \bar{v}_{\rm Hxc}}{\de N}},
\label{GT2_suppl}
\ee
see Eq.~(12) of I. In Eq.~(\ref{GT2_suppl}) the quantity $\d N$ is the 
deviation of the  number of electrons of the HOMO from its half-filling 
value 1. In formulas $\d N$ can be written as the difference between 
$(N-{\rm Int}[N])$ and $1$. In Fig.~\ref{suppl_cond} we show $G_{s}$ and $G$ 
versus $v_{g}$ for different values of the broadening parameter $\g$. The 
value $\g=1.8$ meV is the one employed in Fig.4 of I. For small $\g$ we 
clearly see a pattern similar to that of Fig.~3 in I: if $N$ is close to an 
odd number then $G_{s}$ exhibits a Kondo plateau. The dynamical xc correction 
of Eq.~(\ref{GT2_suppl}) suppresses this plateau and yields the correct CB 
pattern.
\begin{figure}[tbp]    
\begin{center}
\includegraphics[width=0.7\textwidth]{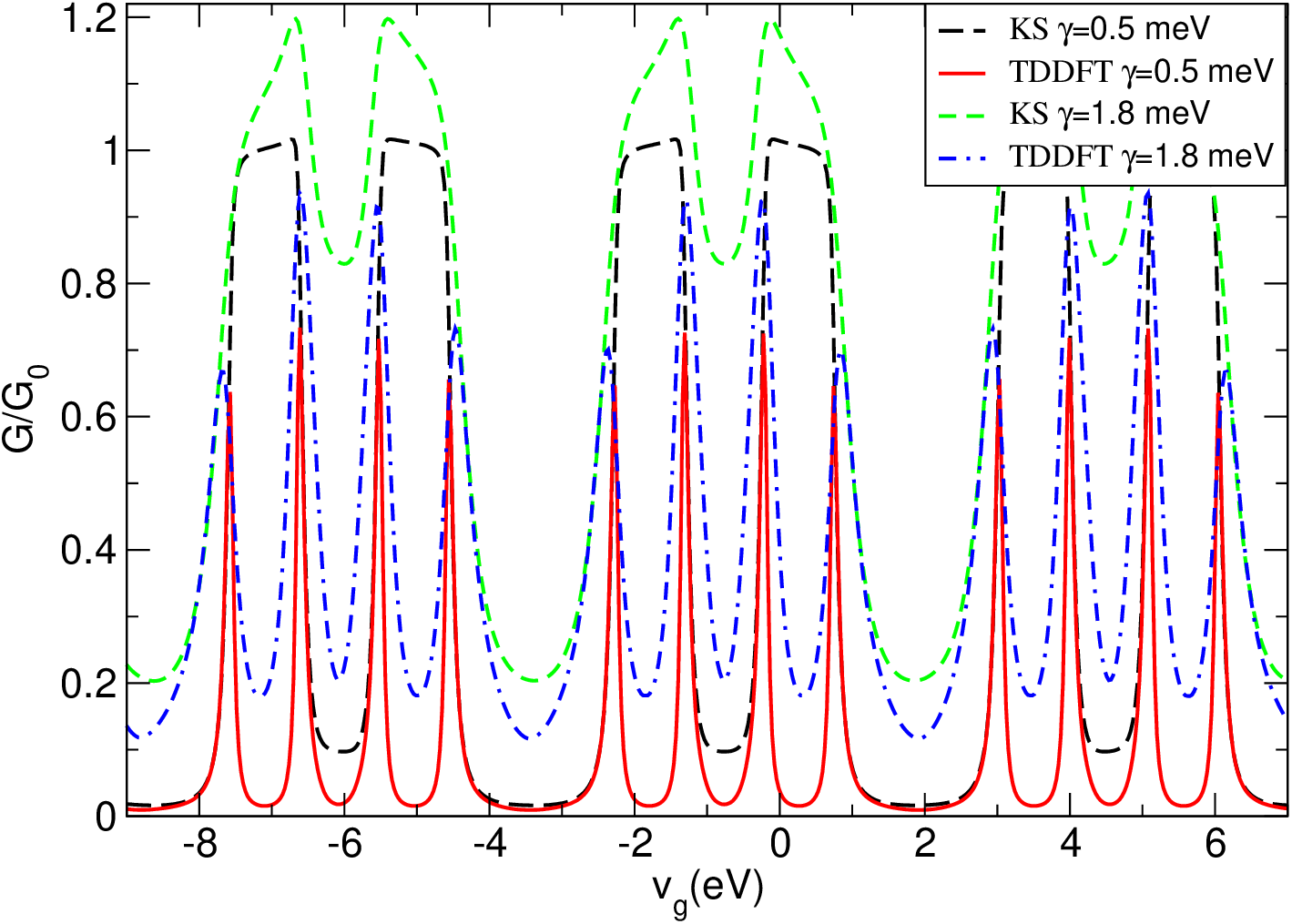}	
\caption{KS and TDDFT corrected conductances for the SWNT quantum dot 
for different values of the coupling $\gamma$ to the leads.}
\label{suppl_cond}
\end{center}
\end{figure}

\end{widetext}

\begin{thebibliography}{99}
    
\bibitem{cuevas}
G. Cuniberti, G. Fagas and. K. Richter,
{\it Introducing Molecular Electronics},
(Springer, Heidelberg, 2005); 
J.C. Cuevas and E. Scheer,
{\it Molecular Electronics: An Introduction to Theory and Experiment}, 
(World Scientific, London, 2010).

\bibitem{sa.2004}
G. Stefanucci and C.-O. Almbladh,
Phys. Rev. B {\bf 69}, 195318 (2004); 
Europhys. Lett. {\bf 67}, 14 (2004).

\bibitem{szvv.2005}
N. Sai, M. Zwolak, G. Vignale, and M. Di Ventra, 
Phys. Rev. Lett. {\bf 94}, 186810 (2005). 

\bibitem{kbe.2006}
M. Koentopp, K. Burke, and F. Evers, 
Phys. Rev. B 73, 121403 (2006).

\bibitem{vv.2009}
G. Vignale and M. Di Ventra, 
Phys. Rev. B {\bf 79}, 014201 (2009). 

\bibitem{rg.1984}
E. Runge and E. K. U. Gross, 
Phys. Rev. Lett. {\bf 52}, 997 (1984).

\bibitem{tfsb.2005}
C. Toher {\em et al.},
Phys. Rev. Lett. {\bf 95}, 146402 (2005).



\bibitem{pplb.1982} 
J. P. Perdew {\em et al.}, 
Phys. Rev. Lett. {\bf 49}, 1691 (1982).



\bibitem{mkns.2010}
H. Mera {\em et al.},
Phys. Rev. B {\bf 81}, 035110 (2010);
H. Mera and Y. M. Niquet,
Phys. Rev. Lett. {\bf 105}, 216408 (2010).
    
\bibitem{sk.2011}
G. Stefanucci and S. Kurth,
Phys. Rev. Lett. {\bf 107}, 216401 (2011).

\bibitem{blb.2012}
J. P. Bergfield, {\em et al.},
Phys. Rev. Lett. {\bf 108}, 066801 (2012).

\bibitem{tse.2012}
P. Tr\"oster, P Schmitteckert and F. Evers,
Phys. Rev. B {\bf 85}, 115409 (2012).

\bibitem{hjbook}
See, e.g., 
H. Haug and A.-P. Jauho, {\em Quantum Kinetics in Transport and Optics of Semiconductors}
(Springer, 2008).



\bibitem{note}
The static DFT kernel is the response of the 
{\em equilibrium} $v_{\rm xc}$  to 
a density variation and, deep inside the leads, is determined  by the condition of charge 
neutrality alone.  

\bibitem{hg.2012}
This follows from the fact that $\d v_{\rm xc}({\bf r},t\ra\inf )$ is defined 
up to the addition of an arbitary constant $C$, see also
M. Hellgren and E. K. U. Gross,
Phys. Rev. A {\bf 85}, 022514 (2012).

\bibitem{uksskvlg.2012}
A.M. Uimonen {\em et al.},
Phys. Rev. B {\bf 84}, 115103 (2011).


\bibitem{check}
The compressibility is weakly 
dependent on temperature up to $T\simeq \g$ \cite{msu2004}. 
Therefore the $v_{\rm Hxc}$ obtained by reverse engineering 
is a good approximation even for $T=0$. At the ph symmetric point 
($N=1$) this approximation gives  $\de v_{\rm Hxc}/\de 
N\sim U^{2}/\g$ \cite{es.2011}. At $T=0$ 
the MB spectral function  for 
$\w\simeq \m$ is dominated by the AS resonance 
$A(\w\simeq \m)\simeq \frac{4T_{\rm K}}{\g}
L_{T_{\rm K}}(\w-\m)$ and thus 
$\int f(\w)\frac{\de A(\w)}{\de \m}\simeq
-\frac{2}{\p\g}$.
For the compressibility  we have from the 
Bethe-Ansatz \cite{wt.1983}
$\kappa=(8\g)/(\p U^{2})\left[1+{\mathcal O}(\g/U)\right]$
from which it follows that also $R$ goes like $(U/\g)^{2}$.

\bibitem{msu2004}
I. Maruyama, N. Shibata and K. N. Ueda,
J. Phys. Soc. Jpn {\bf 73}, 434 (2004).

\bibitem{es.2011} 
F. Evers and P. Schmitteckert, 
Phys. Chem. Chem. Phys. {\bf 13}, 14417 (2011).

\bibitem{wt.1983} 
P. B. Wiegmann and A. M. Tsvelick, 
J. Phys. C: Solid State Phys. {\bf 16}, 2281 (1983).

\bibitem{c.2000}
T. A. Costi, 
Phys. Rev. Lett. {\bf 85}, 1504 (2000).

\bibitem{psc.2012}
E. Perfetto and G. Stefanucci, 
Phys Rev B {\bf 86}, 081409 (2012).

\bibitem{interference}
This amounts to neglecting interference effects, see 
G. Stefanucci {\em et al.}, Phys. Rev. B {\bf 79}, 073406 (2009); T. 
Markussen, R. Stadler and K. S. Thygesen, Nano Lett. {\bf 10}, 4260 
(2010); J. P. Bergfield {\em et al.}, Nano Lett. {\bf 11}, 2759 (2011).

\bibitem{lbp.2002}
W. Liang, M. Bockrath and H. Park,
Phys. Rev. Lett. {\bf 88}, 126801 (2002).

\bibitem{bbnis}
M. R. Buitelaar {\em et al.}, 
Phys. Rev. Lett. {\bf 88}, 156801 (2002).

\bibitem{sjkdkz.2005}
S. Sapmaz {\em et al.},
Phys. Rev. B {\bf 71}, 153402 (2005).

\bibitem{kat.2001}
L. P. Kouwenhoven, D. G. Austing, and S. Tarucha, 
Rep. Prog. Phys. {\bf 64}, 701 (2001).


\bibitem{obh.2000}
Y. Oreg, K. Byczuk and B. I. Halperin, 
Phys. Rev. Lett. 85, 365 (2000).

\bibitem{suppl}
see Supplemental Material for details on the implementation. 

\bibitem{discpot}
N. A. Lima {\em et al.}, 
Europhys. Lett. {\bf 60}, 601 (2002); 
N. A. Lima {\em et al.}, 
Phys. Rev. Lett. {\bf 90}, 146402 (2003);
P. Mori-Sanchez, A. J. Cohen, and W. T. Yang, 
Phys. Rev. Lett. {\bf 102}, 066403 (2009);
F. Malet and P Gori-Giorgi, 
Phys. Rev. Lett. {\bf 109}, 246402 (2012);
X. Gao {\em et al.},
Phys. Rev. B {\bf 86}, 235139 (2012);
J. Lorenzana, Z.-J. Ying, and V. Brosco,
Phys. Rev. B {\bf 86}, 075131 (2012); 
E. Kraisler and L. Kronik, Phys. Rev. Lett. {\bf 110}, 126403 (2013).

\bibitem{tddft_xc} 
C. Verdozzi, 
Phys. Rev. Lett. {\bf 101}, 166401 (2008);
S. Kurth {\em et al.},
Phys. Rev. Lett. {\bf 104}, 236801 (2010);
D. Hofmann and S. K\"ummel, 
Phys. Rev. B {\bf 86}, 201109(R) (2012);
P. Elliott {\em et al.},
Phys. Rev. Lett. {\bf 109}, 266404 (2012);
S.E.B. Nielsen, M. Ruggenthaler and R. van Leeuwen,
EPL {\bf 101}, 33001 (2013). 
\end{thebibliography}
\end{document}